# Digital financial services and open banking innovation: are banks becoming invisible?

*Valeria Stefanelli[1], Francesco Manta[1], Pierluigi Toma[1]*

[1]*University of Salento, Department of Economic Sciences, Campus Ecotekne, SP6 Lecce-Monteroni, 73100 – Lecce, Italy*

**Abstract**

Digitalization in many economic sectors drove the financial system to adapt to new paradigms of technological transformation. Moreover, the extant regulatory framework forced the financial system to reconsider its business models and its relationship with the market. Such reasons generated also in the banking sector a new model of competition within the ecosystem never seen before in this sector. The new ecosystem of banks and financial institutions lacks a common framework that not only synthesizes the development lines of open innovation in the banking sector, but also regarding the planification of strategic choices and organisation within the new ecosystems. The present study aims to inquire the strategic positioning of European banks toward their digital transformation strategies, by analysing the decision-making processes that occurred between 2015 and 2019. A qualitative analysis on partnerships and the adoption of Application Programming Interfaces (APIs) development in support of new service models was carried out. Results have relevant policy implications for regulators, linked to the business evolution and the risks of outsourcing, and managerial implications for the followers, specifically on the plan of service integration to diversify and boost their activities in the segment of customer relationship management and care, providing a better user experience.

## 1. INTRODUCTION

In recent years, new digital technologies (machine learning, Internet of Things, big data, blockchain, etc.) have generated profound changes in the various economic sectors at an international level (Downes & Nunes, 2014). Such demand for evolution boost research to follow up new frameworks in the field of innovation management (Chae, 2019), fostering the concept of 'digital innovation ecosystem', stressing the outstanding relevance gained by heterogeneous factors that contribute to the evolution of such networks (Hu et al., 2016). This evolutionary trend, which involve many actors at different levels, has is pivot in the essential role of big data and analytics, which are considered disruptive and transformative elements for innovation (Raguseo, 2018). They, indeed, proved to carry many benefits to the economic activity, determining a considerable increase in investments in big data skills and resources (Gupta et al., 2018). The changing pattern of the new digital ecosystem is forcing the banking sector to adapt to new business models that take into account the need for digitization and the rethink of their core services and processes to better dialogue with customers (Bank for International Settlements, 2018). This process is both spontaneous, pushed by the birth of financial technology, and fostered by governments in the frame of the EU digital agenda, which require financial firms to encourage the switch to digitization and the adaptation of users to digital products and services. In some areas, indeed, the process of open banking development is market driven (i.e., China and the US), in other countries is prescriptive (i.e., the EU) or in other is encouraged by the issues of guidance and central support (Bank for International Settlements, 2019). The financial sector is among those most affected by this change. Traditional banks, in fact, are committed to repositioning their business model, both for the need to recover efficiency and profitability and to seize the opportunities offered by digital technologies. In this glance, the shift to the business model of Banking-as-a-platform (Omarini, 2018) is shaping the new paradigm of sustainable and attractive opportunities for financial institutions. These changes are also necessary to respond to changes outside the sector, linked to the evolutions and new needs of increasingly digital and demanding consumers (Financial Stability Board, 2019; Bank for International Settlements, 2020; Bank of Italy, 2021).

Such changes also include recent regulatory interventions, like the PSD2 legislation on digital payments, aimed at regulating competition access to the European payments market and guaranteeing greater protection and transparency for users (Associazione Bancaria Italiana, 2020). In this perspective, the PSD2 has recognized the access to the market of new operators, the so-called Fintechs, whose market competitiveness lies in high technology, used for the supply of innovative products or services, in the ability to analyze large amounts of data, in the dynamic and lean business structure. Since these are very often start-ups, these service providers are subject to a preliminary regulatory framework still in definition, instead of that envisaged for credit institutions, which are subject to stricter norms. These opportunities have allowed Fintechs to undermine certain points in the value chain of banking services, focusing mainly on the payments sector, but generally having a widespread throughout the financial services sector. Thus, many differences at different territorial levels, does exist in the state of the art and the evolution of the Fintech sector (International Monetary Fund, 2019).

Considering this, banks must rethink an internal strategic and organizational reorganization in order not to risk losing significant market shares. Regardless of the strategy that each bank intends to adopt, the digital transformation of the business has now become inescapable (Omarini, 2018; Anand & Mantrala, 2019; Khanboubi & Boulmakoul, 2019; Ayadi et al., 2021; Appio et al., 2021). According to a report issued by the Italian Bank Association (ABI - 2022), banks are constantly converting to a service provision mix more and more oriented toward the digitization of their core products: the main strategy pillars watch to the reduction of physical desks, the increase of services provided via ATM, and the increase of interfaces for customers, like the increment of POS payment devices provided and the sharp increase of electronic payment transactions (European Commission, 2015; European Banking Authority, 2018). Some scholars argue that such changes are becoming more effective as banks are intensifying partnerships with Fintechs and hiring digital officers (Hornuf et al., 2021) All these changes have been enhanced by the breakout of the pandemic, which reduced and resized the physical activity of financial intermediaries and enhanced the attitude of consumers toward digital and mobile retail banking (Baicu et al., 2020).

Some studies (Temelkov, 2018; Vives, 2019), underline how Fintechs can qualify as an opportunity for banks that intend to enter productive and/or commercial collaboration and partnerships, in a logic of open banking. Banks, by sharing their data with external fintech or big tech subjects (such as GAFAAs), offer new services and financial products (or already existing services, but provided in a more effective and efficient way), which constitute an added value for customers. Everything is provided through platforms open to collaboration and integration through interoperability standards between different software, i.e., the Application Programming Interfaces (APIs, Premchand & Choudhry, 2018).

In light of this context, the aim of the present study is to examine the characteristics of the banking ecosystems, which, according to our best knowledge, are barely explored. For this reason, this work analyzed the strategic positioning of the major European banks in terms of digital transformation, observing the choices made by those financial institutions in the period between 2015 and 2019. The assessment was conducted by taking as a reference the partnership choices made with Fintech and API development initiatives. The analysis conducted made it possible to frame the strategic direction of the banks under investigation, as potential followers of the overall banking system, considering the appropriate changes and timing. The results of the analysis have a significant impact on banking policies and supervisors, financial intermediaries, and managers, Fintechs and start-uppers.

The remainder of the paper is organized as follows: paragraph 2 outlines the theoretical framework and studies on the topic of digital banking strategy, to support the formulation of the research question. Paragraph 3 describes the methodology, sample and data adopted in the empirical analysis to support the construction of the strategic positioning map. Paragraph 4 illustrates and discusses the results. The last paragraph illustrates the conclusions of the work, the limits and possible policy and managerial implications for the banking sector.

## 2. BACKGROUND AND LITERATURE REVIEW

### 2.1 Digital and financial ecosystems

In the last two decades, there has been a growing interest in the concept of ecosystem in managerial literature and practice (Iansiti & Levien, 2004; Dhanaraj & Parkhe, 2006). The financial sector also highlights the emergence of digital ecosystems (Palmiè et al., 2020) based on financial technology (Fintech). In this work, the emphasis is on the concept of disruptive innovations and how this can destroy existing industries. In this context of disruptive and large-scale evolution of digital economy, banks and financial institutions are the most involved in this running pace evolutionary path, because of their central role of intermediation among actors, as well as the products and services provision to firms and private entities. Intermediation costs, payment services, loans and trading are of utmost importance in assessing the efficiency of all the economic sectors, and the creation of a highly competitive environment in terms of time and financial costs require the configuration of new business models. Moreover, together with public spending, banks and the financial system are the most important actors supporting the economy in the processes of green and digital transitions (Bank of Italy, 2022).

Fintechs have evolved rapidly and are reshaping the banking sector, payments, commerce, financial investments and even the very concept of money with digital currencies. The FinTech ecosystem includes both established companies such as banks and insurance companies, and completely new companies (such as start-ups) that develop new concepts and new business models based on Fintechs. Unsal et al. (2020) also provide a technological approach, based on APIs, to build a financial ecosystem.

Although Fintech is a sector characterized by strong growth - especially in the last 5-6 years -, it represents only a small market share compared to financial institutions (Vives, 2017). This growth was most evident in the US and China, with Europe slightly lagging behind. The United States has the largest Fintech industry in the world with 4.7 million companies. Most of the European Fintechs are concentrated in the UK with 820,000 companies (Riyanto et al., 2018). They have mostly developed in the insurance sector, deposits, loans, and capital raising, digital payments, investment management and financial advisory (Financial Stability Board, 2017). For all these market segments, Fintechs have the potential to reduce brokerage costs and expand access to finance for customers not served by traditional financial institutions.

For banks, one aspect to consider is that one regarding Big Data management. Financial institutions are currently facing the industry transition to Big Data (or the open banking platform paradigm) and for this reason, according to the literature, there is a link between the financialization of the economy and a significant increase in information flows in the banking sector (Klioutchnikov, et al., 2019).

Moreover, Vives (2019) describes digital disruption in the banking sector by examining its impact on competition and its potential to increase the efficiency and well-being of consumers. The results highlight the ability of disruptive new technologies to lead to greater efficiency in the provision of banking services.

Therefore, banks will move towards a model based on a customer-centric platform, which will be managed through the Application Programming Interface. These changes represent major challenges for incumbents, who must increasingly seek to reduce the costs of physical branches and try to reach new standards of services offered, to face the strong external competition from new entrants.

Whether Fintechs represent a threat or an opportunity for incumbents will largely depend on the bank's strategic approach and its desire for cooperation (Temelkov, 2018). The difficulties for SMEs to receive loans from traditional banks, together with the other factors listed above, have made the breeding ground for fintech start-ups for every service offered by the traditional bank (Patwardhan, 2018). In this situation, banks risk losing customer base and profits. According to the literature, due to competition, banks risk losing approximately between 29% and 35% of their revenues (European Banking Authority, 2018). But if banks started the digital transformation process of their business in time, they could not only maintain their position as industry leaders, but they could also increase their profits. This potential loss of profits is already in itself a valid reason for banks to consider the prospect of collaborating with fintech companies. This aspect is of high concern from a supervision perspective: according to the Eurosystem supervisory body, the digital transformation and the cybersecurity of financial institutions are respectively in the priority axes 2 and 3 for the next three years (European Central Bank, 2022).

## 2.2 The digital challenge for banks

In addition to the same analysis, a study conducted by the consulting firm Accenture with 25 senior innovation managers of the participating banks within its Innovation Lab in London and Dublin, highlights three issues to be addressed for digital transformation: the degree of openness, collaboration, and investment (Skan et al., 2014). In the first case we refer to a process, on the part of large financial organizations, of involving external technological solutions. This is widely present in the fintech approach culture and translates into the integration of external systems with the banking core business using specific technologies, the so-called APIs, to offer value-added services.

According to the report conducted by Ernst & Young in 2017 (Ernst & Young, 2017), both the actors involved will obtain advantages from the stipulation of commercial partnerships, only if correctly addressed. Leading organizations should seek the simplification of internal processes and increase the use of external utilities, platforms, and micro-services where possible. All with a view to a component-based architecture, which resembles a set of interoperable building blocks, which will drive next generation innovation and efficiency.

Continuing with the analysis of the literature, to give more value to their customers, amid the regulatory framework offered by the PSD2, banks should offer their products and services in an open architecture of Personal Financial Management, offering for example account aggregation services, as well as comparison skills on commissions, rates, and performance across different accounts. Indeed, many financial institutions look to the future of their business as digital platforms with some physical branches, explicating the open banking paradigm (Rousseau, 2019; Camerinelli, 2020).

This paradigm entails a new vision and a new way of operating. Many financial institutions, including in the Private Banking sector, have started hiring managers from Silicon Valley, integrating new tools and functionalities into their core businesses, typical of other sectors of commerce, through APIs (Bender, 2015).

Nazaritehrani and Mashall (2020) highlight the positive correlation between market share and profit for banks, statistically analyzing this relationship through a survey conducted at the most important Iranian bank, Shahr Bank in Tehran. In particular, the study reveals the positive relationship between the impact of the use of e-banking channels (considering ATMs, POS, Internet banking, Mobile banking, and Telephone banking) on the growth of profit and market share.

*2.2.1 Open banking innovation*

By focusing on the meaning of open innovation, literature identifies two main types of banking strategies: inbound or outbound strategies (Huizingh, 2011). While the latter refer to innovation activities that leverage technological capabilities beyond company boundaries, the inbound open innovation strategy refers to the internal use of external knowledge (Saebi & Foss, 2015). Banks are trying to maintain their relevance in this rapidly changing landscape through various initiatives. The advent of the Open banking paradigm and the relevant regulations (PSD2) have forced banks to review their business models and undertake partnership strategies with the numerous players in the digital financial ecosystem.

Soloviev et al. (2018), analyzing the banking and fintech landscape in Russia, show that the attitude of traditional banks and financial companies is to develop fintech initiatives internally, through collaboration between different entities that are part of the financial ecosystem. Despite this, the main changes for banks concern the use of new technologies, such as artificial intelligence or machine learning, to mainly reduce the costs of current business processes and not to develop new products or services.

Lee et al. (2018) describes the main challenges that banks will have to face in the coming years: i) the correct management of investments/partnerships, choosing whether to invest internally in fintech projects by competing with start-ups or whether to invest directly in these fintech start-ups to stay at the forefront of technology without requiring a total internal renovation; ii) customer management, who are increasingly digital and demanding; iii) the management of regulation, which can be very costly for traditional financial institutions; iv) the challenge of technological integration with existing legacy banking systems, essential for offering the same experience to the consumer on different channels; v) the challenge of security and management of sensitive customer data.

According to the extant literature, banks' preferred partnership with Fintechs is a ready-to-use white-labelled solution. In this case, the financial services company buys a ready-made solution from a fintech and implements it under its own brand, in order to reduce time-to-market.

Another preferred collaboration model is that of integrated in-house solutions. Here, the products and solutions are hosted internally - typically for larger companies - or as software-as-a-service for smaller companies. The financial services industry, therefore, will never be the same again. The ecosystem will become increasingly competitive and efficient by offering consumers much more choice between different players (Soloviev et al., 2018). To be able to maintain their leadership, incumbents should adopt the right mix between an internal technology structure based on an API managed platform and a portfolio of collaborations with innovative external partners that guarantees the development of new products or services with greater added value for the end customer.

**2.3. Purpose of the research**

In the present work, in the light of the literature analyzed, exploratory research was carried out on two main dimensions, to map the directions of development of the digital strategies of European banks. The study has the purpose to open a new strand looking at the birth and development of financial ecosystems, which are recent and still little explored, according to our best knowledge. This is represented by the quantitative mapping of API technologies and the number of strategic partnerships entered by the institutes under analysis with Fintech entities. It was decided to analyze the number of APIs, as the Open

Banking model is based on the construction of an architecture consisting of blocks and managed via APIs, which allows the integration of multiple players on the market. This mapping made it possible to understand which institutes are already adopting this new paradigm and which ones are adapting, proposing the minimum number of APIs required by the PSD2 regulation and what strategic partnerships they use. With reference to these collaborations with Fintech players, as already repeated in the literature, it is an essential factor for the outsourcing of some banking services, in order to acquire the necessary technological skills, enter markets adjacent to the banking one, develop new services to greater added value, making current business processes more efficient.

In the literature, many studies focus attention on the new Open Banking paradigm, pointing out how relevant is the role of APIs in building such paradigms, but no one considers the number of APIs as a fundamental dimension of the degree of openness and competitiveness of a bank. The search for agreements with strategic Fintech companies is always considered as an essential factor to compete. In all the previously analyzed studies, there is the awareness that collaboration is the key to the Banks-Fintech relationship. On the other hand, the development of APIs as a distinctive and competitive element for all banks is of marginal importance, proposing only guidelines on how the software and organizational architecture of each bank should be structured.

However, according to this analysis, a bank with a high number of APIs demonstrates that it has made significant investments in the past to adapt to technological and market changes. It also demonstrates, in some cases, that it has made the right strategic partnership choices. In addition, banks that have different API functionalities have a high number of commercial agreements entered. In fact, we intended to analyze how the degree of openness of a credit institution affects the stipulation of external strategic agreements and partnerships. This was done by analyzing the qualitative aspect, precisely mapping the number of APIs developed by the banks and the number of partnerships found from 2015 to 2019. Choosing such variables to measure the orientation of European banks is suggested and supported by several factors. First, PSD2 regulation on digital payment systems impose banks to guarantee a minimum number of APIs internally developed. APIs are considered an essential element to pursue Open banking objectives (Premchand & Choudhry, 2018). Second, outsourcing of services, in this transition phase, are essential for incumbent banks to keep the pace of Fintechs, which are gaining a greater importance in the sector (Juengerkes, 2016). On the other hand, scholars and practitioners argue that not all the financial services can be translated into fully digital interfaces (Vatolkina et al., 2020). For this reason, it is appropriate to inquire also which is the extent of evolution of digital financial services according to the market demand.

We formulated two research questions to centre the focus of our study on the dimensions object of analysis:

*RQ1: to which degree European banks developed internal API functionalities?*

*RQ2: to which degree European banks promoted external partnerships with Fintechs?*

*RQ3: to what extent the financial ecosystem is pursuing the demand of the market?*

From the intersection of the two dimensions, a possible strategic reference framework is obtained, consisting of four quadrants, within which to place the analyzed institutions. This made it possible to frame the strategic direction of the banks under investigation. The reason standing behind the present analysis is the outstanding penetration of Fintechs among European consumers, with peaks of more than 70% among users in some countries (i.e., The Netherlands and the UK), with growth pace rates of more than 400% since 2014 to 2019 (Ernst & Young, 2019). This incredible growth is not uniform among countries, but the only element of certainty is related to the typology of products and services, which are gradually shifting the focus from only payment-related applications to a broader range of banking, insurance, asset management and personal finance apps. This broad expansion of digital shifting, under the umbrella of PSD2 regulatory, banks are endowed with the duty of developing open APIs and are willing to promote partnerships with Fintechs on many aspects, especially data sharing to design tailor-made services for customers.

## 3. METHODOLOGY, SAMPLE AND DATA

To answer our research questions, the analysis sample was first identified. The largest European banks were identified on the basis of total assets, according to data provided by S&P (2019). The sample consists of 18 European banks, also considering the United Kingdom (UK) within the perimeter of analysis. In order to adequately diversify the sample, the top three banks for the main European countries were taken into consideration: Italy, Germany, Spain, France, Holland, UK.

To collect information on banks, given the absence of official and predefined databases, a qualitative analysis methodology was adopted, based on the collection of articles from newspapers, scientific journals, found through the internet by choosing appropriate keywords. In addition, reports, results of specific surveys on the topic of collaboration as a key success factor for the financial sector were consulted, to have a clear picture of the modus operandi of credit institutions in the digital era for the period 2015-2019. The choice of the considered timespan is due to the disruption of the phenomenon in its manifested and actual features. Indeed, although the Fintech sector has been growing since 2009, at the tail of the global financial crisis, it is only in 2014 that massive investments in the sector has started to flow throughout the European financial market. First regulatory were issued in 2015 (PSD2), but came into force only in 2017, so that the phenomenon had a natural evolution to 2019.

With reference to the two dimensions of strategic positioning observed, the number of APIs developed by the bank and the number of partnerships created, specific sources were used. In the case of the first, the number of APIs developed by the bank, information was found on the online portals of each institution, within which a page dedicated to third-party developers was always found (except for Bank K) parts and relative PSD2. On each page, the functions that can be performed and integrated by third parties through different APIs together with the same bank are listed. In addition, all the banks analyzed have virtual spaces, called *sandboxes*, within which developers can, with the authorization of the competent authority, test their codes and ideas in a completely safe manner (Alaassar et al., 2021). Regulatory sandboxes are widespread all over the world and in the European Union and have been enabled with the scope of testing innovation especially in the Fintech sector. In the EU the Netherlands have activated a functioning live sandbox, so has done the Bank of Italy too, which launched a call for firms to join the latest pilot project on implementing innovation in the financial and insurance sectors (Bank of Italy, 2021). It should be noted that the number of APIs adopted by each bank is not an exhaustive indicator of a high internal technological specialization but, given the advantages that the APIs themselves have, this can represent a distinctive and advantageous / competitive factor towards the market. and its banking and fintech competitors. The exploratory research on the number of APIs has made it possible to obtain a further classification of the APIs, based on the functionality that these codes offer. This is illustrated in Figure 1, indicative of the number of APIs for each bank and their respective financial services capabilities.

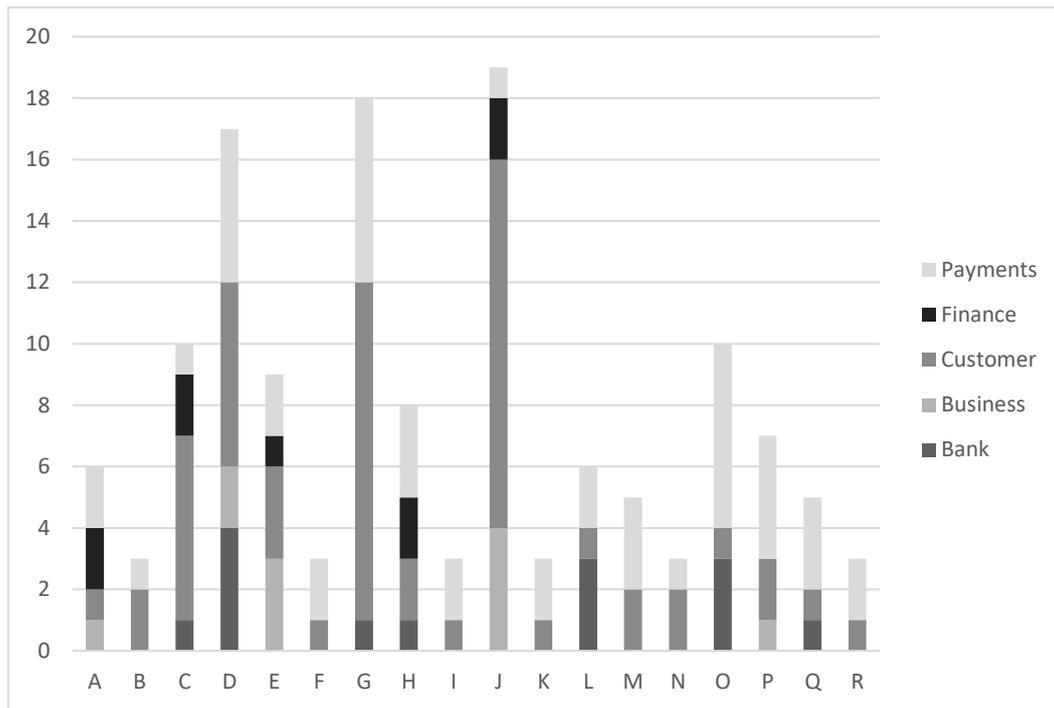

*Graph 1 – APIs: number and functionalities (Source: own elaboration on data found on banks' websites)*

What emerges from this graph is certainly the regulatory reception of the PSD2, considering that the APIs most used and developed by all the banks in the sample are those corresponding to the Payments and Customers services, field of application of the PISPs and AISPs introduced. Institutions with a greater number of APIs assume a greater and certainly earlier predisposition than competitors to the open banking paradigm. Banks that previously invested in technological, infrastructure and organizational changes are now one step ahead. This strategy is a call for institutions to fulfil the transition to Banking-as-a-platform business models (Omarini, 2018).

As regards the second dimension of the strategic positioning matrix, the number of commercial partnerships entered with fintech entities, the information was found at a site for research and aggregation of corporate economic-financial data, namely Owler.com. Along with the latter, the website of another research and data aggregation company, Crunchbase.com, was also consulted.

Investment transactions (mostly in Equity, but also in Debt and in the various life stages of the invested start-up) and acquisitions, where existing, of fintech companies were considered commercial partnerships. To further extend the sample, it was decided not to include only the agreements vs the real fintechs, but to also consider those companies equipped with innovative technologies, related or inherent to the banking and financial sector (i.e., start-up that deals with creating graphical interface for apps).

Furthermore, the transactions listed above were taken into consideration for a period of time from 2015 to 2020. Based on the information collected, graph 2 was created, which illustrates the number of agreements entered into by the sample with external partners.

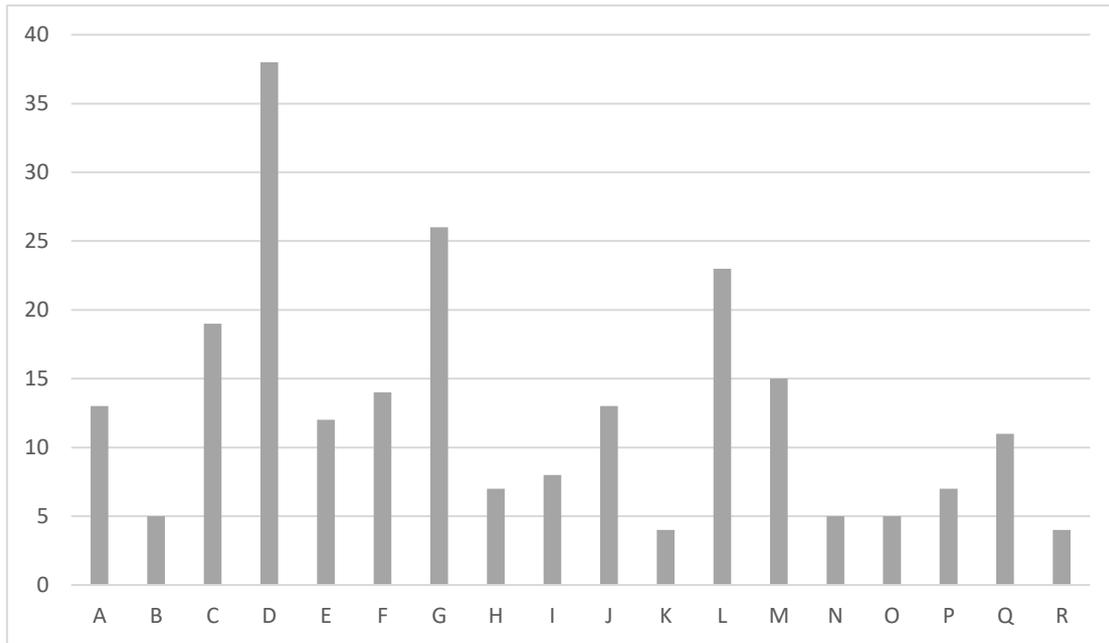

*Graph 2 - Number of partnerships signed by the sample of banks in the 2015-2020 period. Source: own elaboration on Owler.com data (2020).*

Bank D dominates the panorama with more than 35 partnerships signed, with H and L following respectively with 26 and 23 agreements made.

The two dimensions analyzed shall be considered quite correlated, but in any case, it can be said that a higher degree of technological specialization and openness, thanks to the presence of numerous and various APIs, could positively influence the stipulation of partnerships with innovative and external fintech entities to the company. Despite this, in the same analysis, there are cases in which a low number of APIs (equal to 3, as required by PSD2 regulation) corresponds to a high number of collaborations and strategic partnerships. Much also depends on corporate strategic plans, which in most cases involve the adoption of an increasingly digital business model.

**4. RESULTS**

What emerges from this quantitative mapping of the banks examined is the following graph 3, called the "map of strategic digital innovation choices of European incumbents". By placing the two quantitative dimensions obtained on a Cartesian plane, it is possible to position the analyzed banks in certain points of the plan.

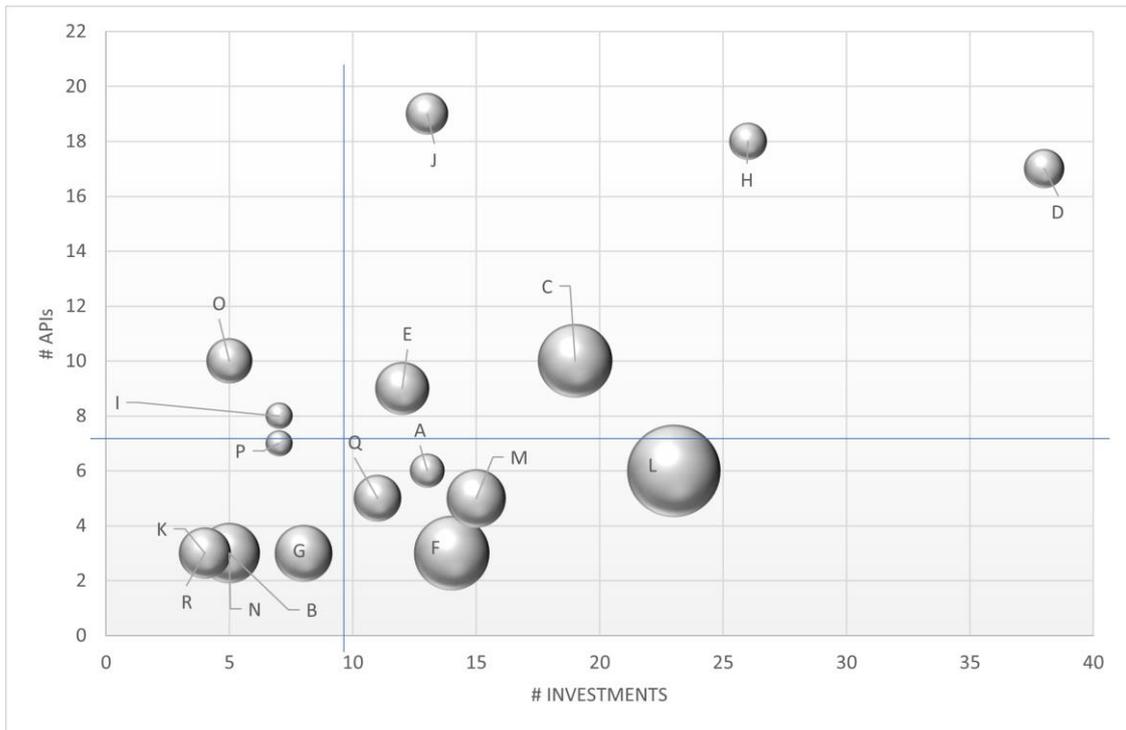

*Graph 3: Map of the strategic digital innovation choices of European incumbents (Source: authors' own elaboration).*

Before moving on to the description of the results obtained, it is worth making some observations:

- all the banks analyzed have included development programs based on technology and innovation within their business plans;
- some banks have already been emphasizing digital development projects for some years more than other European competitors;
- all banks are to be considered aligned with PSD2 regulations, offering services via API to enable payments to PISPs, or aggregating current accounts to AISPs or checking the availability of funds against a payment requested by CISPs;
- all banks offer third-party developers a "sandbox" environment, which is necessary in order to test the services they want to offer;
- the two dimensions, as previously mentioned, are correlated, but by integrating them, it is possible to declare, at least for the analyzed sample, which banks are currently in a competitive advantage over the others and over FinTech competitors;
- the research and analysis conducted should not be considered exhaustive with regard to the European banking landscape but can provide useful information on the respective competitors not mentioned in the analysis, in order to trace a trend more or less followed by most banks financial.

The arrangement of the banks analyzed in different quadrants of the graph allows to identify four strategic competitive behaviours on European banking market, represented in a crescent order:

- the first group, consisting of those that have been defined as "Follower" banks, i.e., are those banks that have adapted to the regulatory changes envisaged by proposing the minimum number of APIs necessary to comply with the legislation and boast a low number of collaborative investments towards strategic external partners; here we find all the three Italian banks, a German bank and a French one. In this case, results show to be in line with the cited parameters: indeed, Italy ranks among the least digitalised countries, as well as it ranks among the worst countries for digital payments transactions per capita, so as France do;

- the second group is represented by the banks called "Innovative", because they are institutions that prefer to outsource the digitalization process, at least in part, by entering important partnerships with FinTechs specialized in targeted market segments; despite this, these banks also have a good technological level, as it is possible to find some additional services offered via API. In this frame we find two Dutch banks, two French banks and a British one. In this quadrant we find those banks where an effort to internally implement APIs is at an early stage, but some of those banks are slowly adapting to the needs of their users, which (in the case of Dutch and British) are more demanding for service dematerialization.
- the third group, represented by the "Technological hubs" banks, i.e., are banks that not only comply with the terms of the legislation by proposing standard APIs, but even offer additional ones, with different and varied functions which obviously represent services with greater added value. Being highly technologically developed, these players do not exclusively focus their attention on external commercial agreements, which in any case are present in smaller quantities; here we find a German, a Dutch and a British bank. According to what stated before, it is only to add that Dutch people are even more digitally educated than the other two countries and are also among the countries where digital payments per capita are far higher than the rest of Europe. This could turn out to be a relevant issue for Dutch institutes, which ought to be in line with the demand for services of citizens and users;
- the last group is represented by the most advanced banks, those that have competitive advantages due to a strong presence of internal technological skills and competences, but also thanks to strategic agreements and collaborations with very important strategic partners; they are called "Invisible banks". Those are banks C, D, E, H and J; three of them are Spanish, one British and one German. This data is quite intuitive if compared with parameters of digitization and digital literacy of the citizens of those countries: indeed, all those three countries have a DESI Index (Digital Economy and Social Index) far above the EU average. Data about digital payment transactions per capita is in countertendency, showing how both Germany and Spain are under the EU average, while it is well known how in the UK digital transactions already overcome physical payments;

The graphic reported below represents what has just been said.

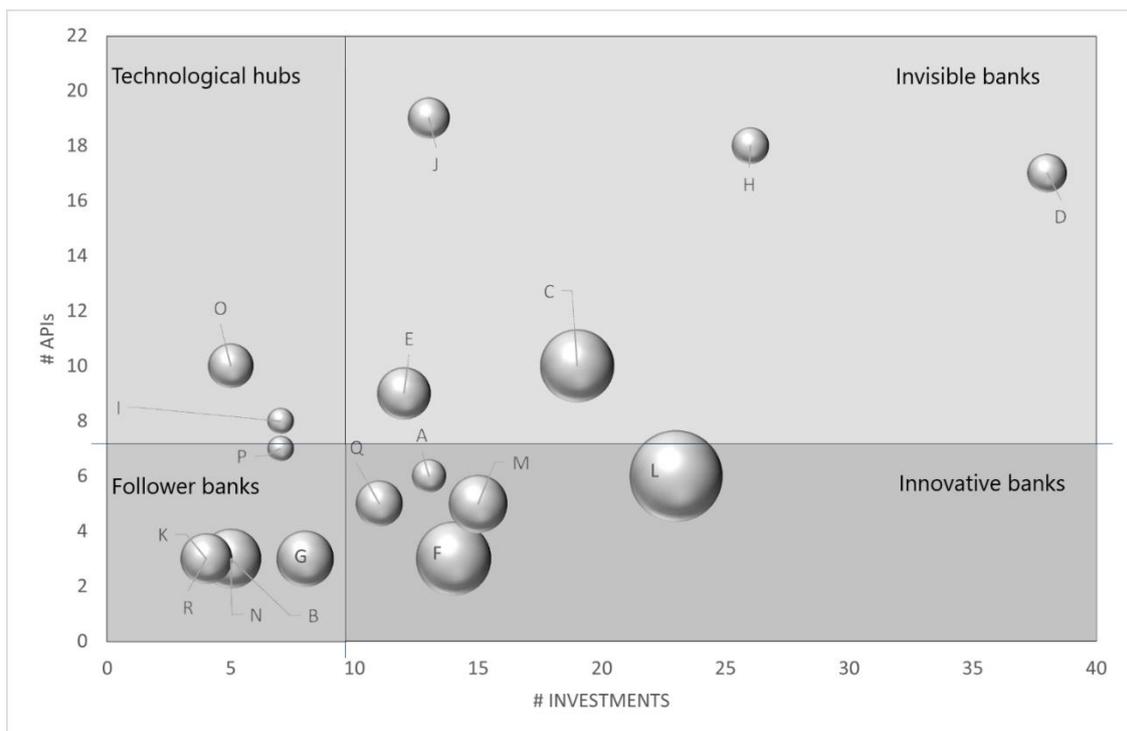

*Graph 4: Reference framework (Source: authors' own elaboration).*

At this point it is possible to make observations on the strategic choices of these incumbents in the digital age. Surely, we can define the Best in Class, the best banks from an internal technological point of view and openness to new and strategic partners, those belonging to the third quadrant, called Invisible Banks. Within this cluster, we find the banks C, D, E, H and J. These institutes have a greater propensity for technological and innovative changes in the market, they do not adopt a reactive approach to legislation, but proactive, riding the wave of innovation and trying to anticipate trends and the evolution of regulations themselves. It is therefore possible to declare that approximately 28% of the banks present in the analyzed sample belongs to this category.

Second, we note a particularity regarding the second quadrant, that of technological hubs, where we find 16% of the sample analyzed, represented by three banks, I, O, and P. This is because it is the result of a management choice that aims to enhance the internal resources of the company, developing a greater number of APIs with different functionalities. Furthermore, the ease of viewing these services on your website should not be underestimated, as they allow a clear understanding of the features offered even to the less experienced of the subject.

In the fourth quadrant, that of innovative banks, we note the presence of three banks, corresponding to approximately 28% of the sample analyzed. In this quadrant we observe the presence of banks A, F, L, M, and Q, characterized by a high number of investments and fewer APIs offered. Despite the positioning in this quadrant, the features offered via API in an Open Banking perspective are different and some are in the testing phase, as in the case of A and the APIs developed to request loans through a procedure using the app, exclusively digital. They are also characterized by a high number of commercial agreements with external strategic partners.

In the quadrant of follower banks, we find the remaining 28% of the sample analyzed, represented by the B, G, K, N, and R. These banks, compared to the remainder of the sample, adopt a reactive approach to the legislation, that is, they possess the requisites required by the authorities, but are unable to adopt a proactive approach in the face of market changes. Such resulting behavioural strategies are coherent with what is argued by Khanna (2018), by expressing the opportunity to create a balanced strategy mix between inhouse developed services and the signature of partnership agreements with third parties.

According to the extant literature, the analysis conducted discloses a landscape in which there is a stringent urgency with respect to the adaptation of the banking system to innovation. Lee et al. (2016) proposed a twofold model of technology advancement. Results tell us how the "outside-in" scheme is largely preferred by the banks included in the studied sample. Indeed, the development direction, as suggested by graph 4, is oriented toward the increase of the partnerships with external entities, betting much more on outsourcing know-how externally more than internalising such operations and business models. This is furtherly confirmed by Bagherzadeh et al. (2019), which stated that outside-in open innovation influence innovation performance. The graph confirms what is argued by Skan et al. (2014), considering the necessity for banks to extend their capacity of innovation by increasing the inhouse development of APIs, as well as strengthening the outsourcing activities by incrementing the number of partnerships with innovative start-ups. Another interesting point of discussion is focused on the human resources management. Indeed, the tendency of many institutions to develop inhouse products according to the market demand, will encourage banks to assume staff with a strong background in ICT (Bender, 2015).

**4.1 Possible strategic choices**

The obtained frame shows a very fragmented situation of business development, and different stages of advancement of the digitization process in the banking ecosystem. Differences emerge from a country perspective, marking the importance both of the cultural context and the business environment. The overview on the average customer in Europe tells us how more than the half of the population is sufficiently digitally educated, but only one fifth of the total can say to be well competent on digital issues (European Commission, 2022). Moreover, larger differences are emerging among countries, showing how some of the countries belonging to the sample are ranked far lower than the EU average.

As previously mentioned, these results are in line with the comparison made on the trend of the DESI index of the countries of provenience of the analyzed banks. Banks from countries like Spain, the Netherlands and the UK show a considerable state of advancement in the digitization process of products and services. Spain, in particular, have all its banks in the quadrant of Invisible banks, that are performing really well in the process of dematerialization of their business. Other countries show a hybrid scenario, so the digitization performance is exclusively depending on the management decisions. On the other hand, serious issues ought to be discussed for Italian and French banks, which show to be backwards. French and Italy are, compared to the other countries in the sample, the follower countries, since they present low performance in the DESI Index, so as are two countries with a very low rate of digital payments transactions per capita (European Commission, 2021). It seems that these last two countries are more affectioned to physical banks than others, although changing strategies at a regulatory stage are ongoing. All of these are element of awareness for the supervisory board of the Eurosystem, which fixed the priorities of intervention for the gap 2022-24 on digitization processes and cybersecurity (European Central Bank, 2022).

Although margins of development are possible for every categorization, the most relevant behavioural lines shall be considered starting from the most backward group. Thus, considering the followers quadrant as the starting one, it is possible to imagine some possible strategic directions in future years:

- a first path represented by those banks which, starting from compliance, will subsequently develop, acquiring technological skills internally or externally, extra APIs in different categories and functionalities, in order to be able to definitively embrace the Open API paradigm over the years;
- a second path, in which the follower banks will prefer to focus on their core business, offering a few extra APIs belonging to a specific category based on their business models, focusing more on agreements and collaborations with technologically advanced external partners;
- a third path that may arise is represented by follower ones who decide to invest in their internal resources, in order to reorganize their business model based on APIs developed internally and with a high number of specific features, to re-enter the sector of technological hubs.

**5. CONCLUSIONS, LIMITATIONS AND IMPLICATIONS**

The present study aimed at analysing the framework of the financial ecosystem of European banks in the light of the digitalization process and the challenges that it has to face in order to contrast the raise of Fintechs in the context of innovation in financial products and services provision. We collected data about the use of APIs within banking services and operations, so as the practices of outsourcing of digital services which banks were not able to internalize and develop inhouse.

This gave the opportunity to obtain a framework in which the biggest European result followers of the evolution path of the new ecosystem, facing challenges which are opportunities and threats for the survival of their market share. The raise and the success of many Internet-based financial companies, which do not envisage and undergo specific, strict regulations, are testing traditional financial intermediaries the trial of modernity. The most relevant result obtained from the analysis is the evidence of a business model that can be defined as the "invisible bank". It is endowed with a paradox itself, defining that bank, which is closer to the customer, but is physically disappearing. Indeed, the more the bank develops internally APIs, the more negotiates external partnerships with Fintechs, the more it disappears from the territory, the more is present in the "cloud". This, of course, is an element that generates pros and cons in his affirmation. Invisible banks are surely representing the essence of the sustainability paradigms, reducing their environmental impact, improving their economic sustainability in terms of profit increase and loss reduction, as mentioned throughout the literature (Cillo et al., 2019), and contribute to the social component on the side of financial inclusion, i.e., accessibility to products and services is more widespread due also to cost intermediary cost reduction (Si et al., 2022). A negative aspect is related to a broader limitation on the digital literacy of users. According to the analysis conducted, all the banks from a country belong to the category of follower banks, i.e., the least digitally developed among the institutes in the sample. This is an urgent indicator referred to at the level of digitization of a country and the demand for digital services from citizen/users/customers.

The hypothesis in which the bank offers specific banking services through a single environment managed via APIs is therefore configured. In this environment the customer, once identified and recognized, can take advantage of all the payment services, and aggregate the information of other accounts in his name, or take advantage of further services with greater added value, which could develop persistently in the coming years. In addition, there is also the presence of Big Tech (Google, Amazon, Facebook, Apple), which have already developed exclusive platforms through which to offer services typically offered by banks. In these cases, the latter, especially if they are small, could also join. Finally, we could imagine a development of a banking business model, based on open platforms (i.e., banking-as-a-platform), in which banks would continue to market their financial products and services, perhaps made by external developers or FinTechs.

Possible limitations of this study fall within the reduced dimension of the sample, which anyway includes the biggest banks in Europe. Another element strictly related to the present analysis is the absence of the evaluation of the phenomenon through a time series, which may capture evolutionary paths and, in this glance, the growth and development pace of the financial ecosystem. Fintechs segment is more represented in other areas of the world (i.e., North America and Far-East), and would be ideal to understand how this changing path is moving in areas with a higher market penetration. Future research will also focus on the possibility to assess this adaptation to innovative business models in compliance with the cultural difference among territories.

Theoretical implications arise from the present analysis: the main contributions we intended to give regards the newly created concept of "Invisible bank", which is fostered by the spread of the banking-as-a-platform business model. From this notion new avenues for further research may rise, regarding the assessment of efficiency and productivity of the financial technology sector, based on the gaps filled by the present study. Managerial implications may rise from the analysis conducted. First, a relevant implication, in this glance, is addressed toward the top management of financial institutions, e.g., banks, to best assess their innovation strategies and face the challenge of business innovation. The focus is mostly targeted to the part of the sample recognized as followers, which manifest to be behind schedule. Moreover, banks must be able to focus attention on its customers and their requests, trying to understand what their needs and requirements are, placing itself more and more, not as a simple bank, but as a multiservice platform, in order to earn over time an important competitive advantage. Second, banks will have to consider Fintechs no longer as a pure competitor, but as an integral part of an ecosystem that creates an effective digital experience for the customer. Remaining competitive in such a rapidly evolving context is a challenge that operators in the sector will have to face, adapting their business models to the changes underway. Third, a relevant managerial focus shall be directed to the customer relationship management; consumers are willing to obtain benefits from open banking, demanding lower costs and more customised services. Customer care and relationships shall be addressed in this direction, with a mutualistic and synergic connection between the bank and the customer, especially for what regard security issues and data protection. Fourth, the rise of the banking-as-a-platform model generates virtuous mechanisms aimed at reducing information asymmetry, encouraging the entrance of neobanks (i.e., new-born financial intermediaries) in the market, enhancing competition. Finally, we observe the need to foster policy-making activities to regulate the Fintech market, usually composed by a microcosm of realities which take advantage of grey areas of regulations, insisting on those lacks limitations that are guarantees for customers and the financial stability of the system. One of the most critical issues regards supervisory issues: since invisible banks are shifting to a banking-as-a-platform model, materially disappearing from the market (e.g., less territorial branches), the supervisory control models for banks need to be reviewed in the glance of developing an effective paradigm that ensures the encompassing management of banks. Regulatory sandboxes then, represent a fruitful solution, in which the experimentation of digital innovations allows banks and Fintechs to develop new solutions compliant with the rules and authorities to understand configurations of new activities and related risks to be controlled in the interest of the stability and competitiveness of the banking system. The standardization of determined processes must be a safeguard for the system at an international level, intervening, for example, on tax uniformity and country registration, in order to provide the system with stability, efficiency and security for a correct competition. The homogeneity within the ecosystem is a prerogative of a loyal competition on the market, although the different business structure of traditional financial institutions urges such paths of change for them.